\documentclass[a4paper]{jpconf}
\usepackage{graphicx}
\begin{document}
\title{Status report of the ANTARES experiment}

\author{Y Becherini for the ANTARES Collaboration}

\address{Dipartimento di Fisica, University of Bologna and INFN Sez. Bologna, Viale Berti-Pichat 6/2, 40127 Bologna, Italy}

\ead{Yvonne.Becherini@bo.infn.it}

\begin{abstract}
The ANTARES Collaboration is building an underwater neutrino telescope in the Mediterranean sea. 
The telescope is designed to search for high energy (E $>1$ TeV) galactic and extra-galactic neutrino sources, 
but could also be sensitive to neutrinos originating from the decay of neutralino and exotic particles.
The detector is a 3-dimensional array of photomultipliers located at a depth of 2500 m, 40 km from 
the La Seyne sur Mer shore (near Toulon, France).
During the year 2005 a full scale test line and an instrumented line have been successfully operated. 
In the winter '05-'06 the first full 480 m line will be deployed and connected to the shore station.
\end{abstract}

\section{Detector structure and expected performance}

ANTARES (Astronomy with a Neutrino Telescope and Abyss environmental RESearch)
is a multi-purpose experiment aiming to give scientific results in particle physics, neutrino astrophysics, 
marine biology and geophysics \cite{Proposal}. 
Starting from the winter '05-'06, the Collaboration will install 12 independent 480 m long mooring lines 
holding optical modules (OMs) for the detection of the Cerenkov light emitted by charged relativistic particles 
propagating in sea water. The detector will also be equipped with devices for environmental surveys. 

The lines will be anchored at the sea bottom at about 40 km off the La Seyne sur Mer shore and will be maintained stretched 
by buoys at the top. 
The spacing of the lines varies from 60 to 75 m, a distance which corresponds to the attenuation length of light at the ANTARES site 
\cite{Transm}.
Each line holds $75$ OMs, each composed of a glass sphere with-standing 250 bars containing a 10-inch Hamamatsu photo-multiplier 
(PMT) \cite{OM}.
The OMs are oriented $45^\circ$ downward. In these conditions, the loss of transparency due to biofouling is 
less than 1.5\% per year \cite{Sedim}.
A group of three optical modules and an electronics container (LCM) are positioned in a \textsl{storey}, 
the fundamental detection unit of the detector.
The vertical distance between two adjacent storeys is 14.5 m, 5 storeys form a \textsl{sector} and 5 sectors compose a line.
Digitised signals from each sector are sent to the String Control Module (SCM) located at the Bottom String Socket (BSS), 
and then to the shore by a 42 km long electro-optical cable (MEOC).

The relative position of the detector elements is measured by a high frequency long baseline acoustic
positioning system, and a set of tiltmeter compasses.
The acoustic positioning precision is related to the uncertainty of the sound velocity measurement. 
For this reason instruments continuously measure environmental parameters such as
the temperature, the salinity and the pressure, as well as the sound velocity itself. 
Most of these instruments are installed on the so-called instrumentation line, a special line which is also meant to 
illuminate the full detector for timing calibrations, using Laser and LED beacons.


Monte Carlo simulations of the full detector including light scattering, absorption and background, and the PMTs Transit Time 
Spread (TTS) have shown that the expected angular resolution will be better than $0.3^\circ$ for $E_{\nu}>10$ TeV. 
The expected sensitivity for observing point-like sources in the southern hemisphere 
after one year of full detector operation is given in terms of 
the 90\% confidence muon flux limit of $(3\div6) \times 10^{-16}$ cm$^{-2}$s$^{-1}$ 
for an E$^{-2}$ dependent flux (which is about a factor of two lower than the MACRO limit \cite{MACRO}).

\begin{figure}[t]
\begin{minipage}{13pc}
\includegraphics[width=18pc]{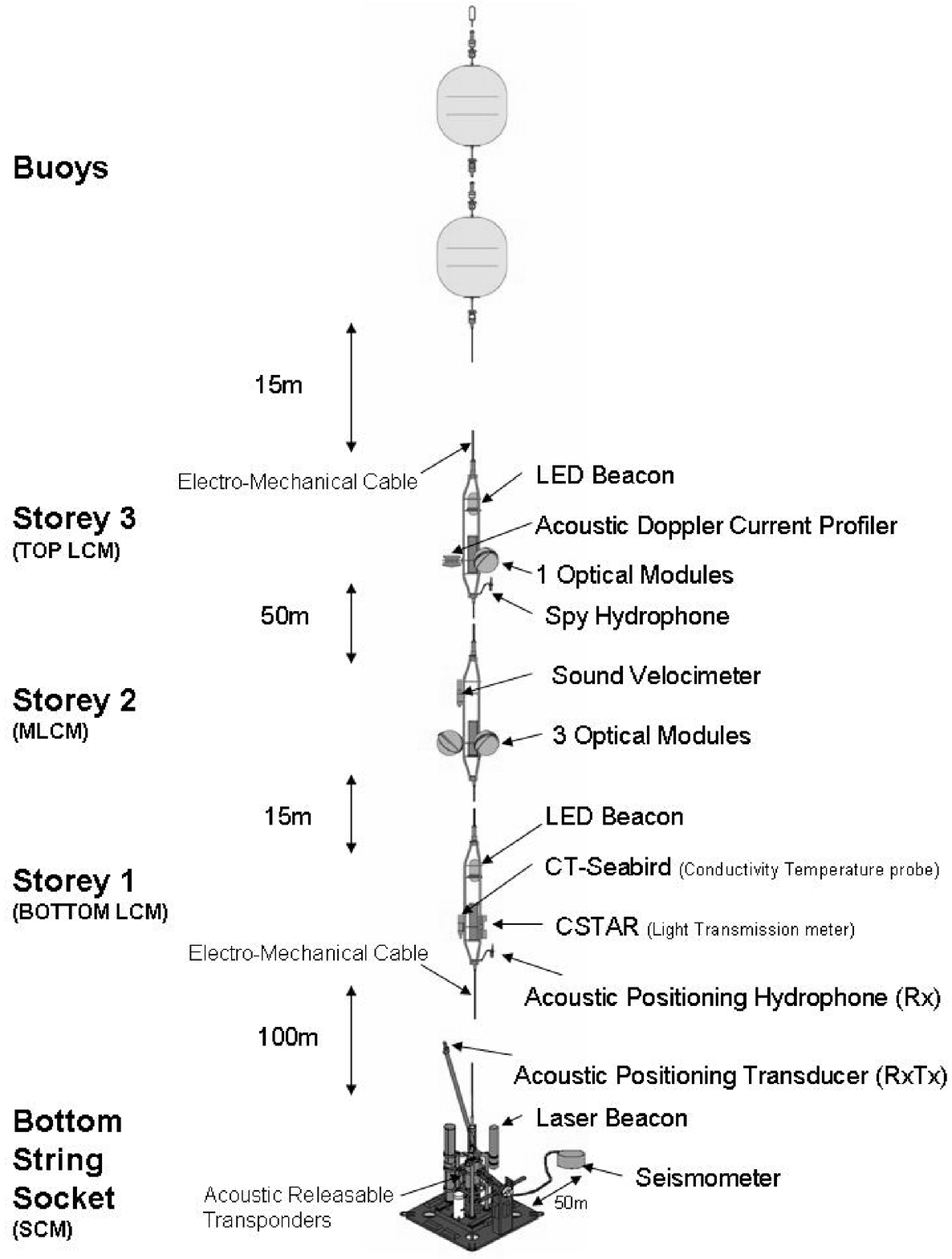}
\caption{\label{milom} The MILOM. The line was deployed in March 2005, was connected in April and it is still taking data 
8 months after deployment.}
\end{minipage} \hspace{5pc}%
\begin{minipage}{20pc}
\begin{center}
\includegraphics[width=17pc]{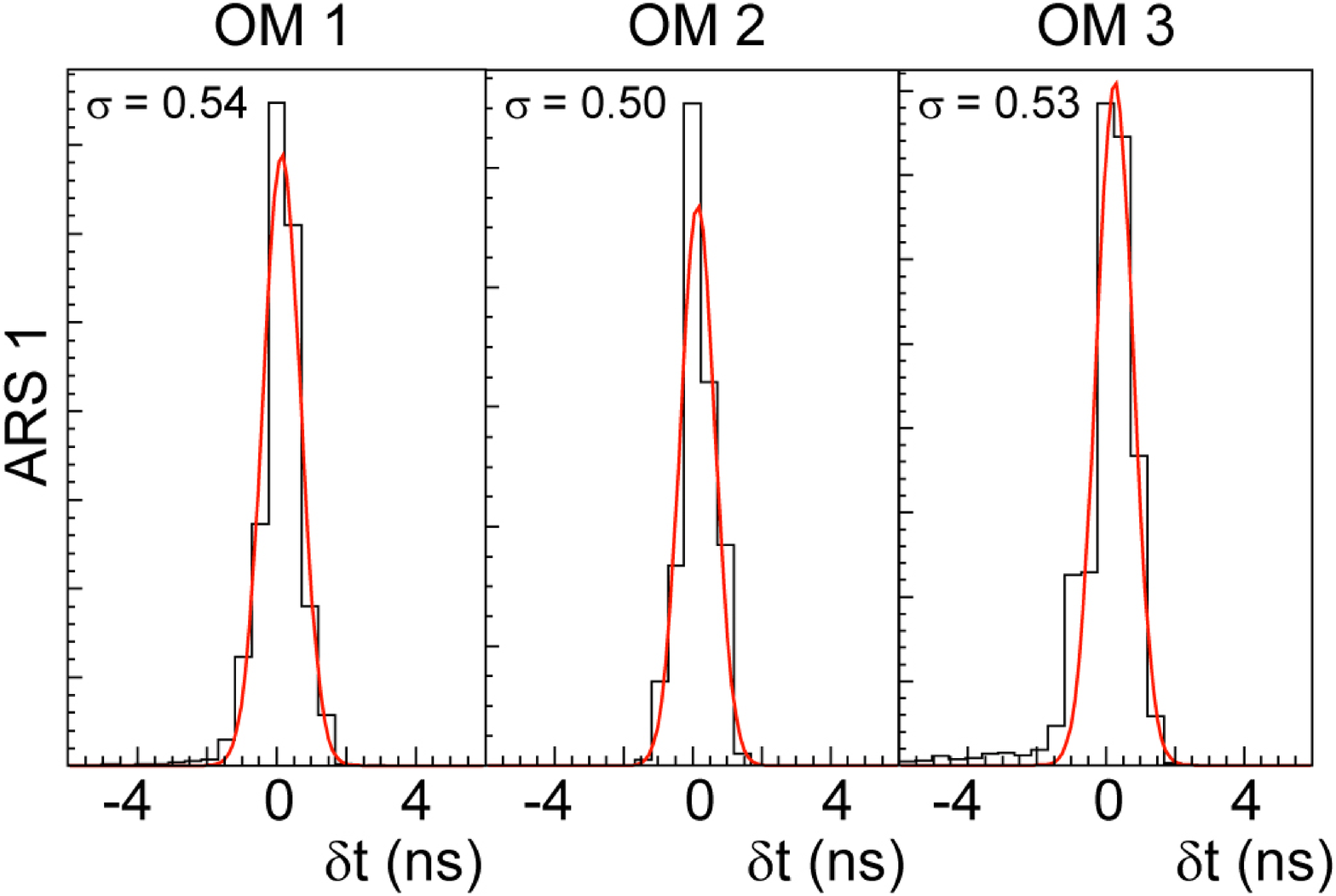}
\includegraphics[width=13pc]{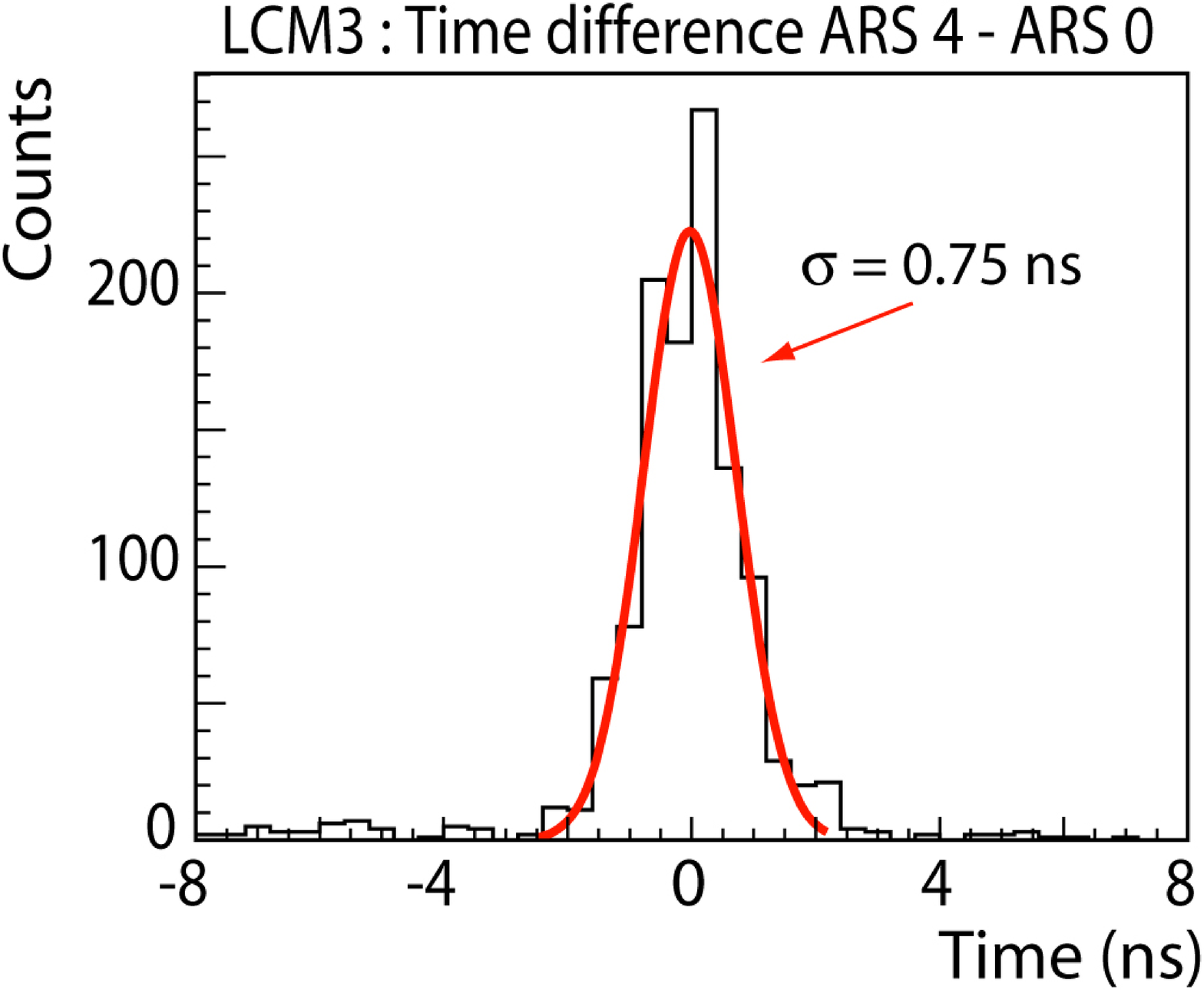}
\caption{\label{data} 
Top: Time distributions of a signal in three optical modules relative to the 
reference signal of the PMT from the LED Beacon after subtraction of 
the offset due to light propagation.
Bottom: Distribution of time differences between two OMs on the MLCM, both flashed by large amplitude pulses from the 
LED beacon located at the Bottom LCM.}
\end{center}
\end{minipage} 
\end{figure}

\section{Line 0 and MILOM operations} 
During the year 2005 the Collaboration has deployed two lines: Line 0 and MILOM. 

Line 0 was the first mechanical test of a line composed by 23 storeys, equipped with water leak sensors 
and instruments for transmission attenuation measurements. 
Line 0 deployment took place in two steps, a first deployment on 15 March 
with the line remaining for 1 hour on the seabed followed by an immediate 
recovery for a fast check, and a second deployment on 17 March with the subsequent connection 
to the Junction Box (JB) by the Remotely Operated Vehicle (ROV) ``Victor'' on 13 April. 
The line was recovered on May 12 and
no water leaks were detected except for a few cubic centimetre in an unconnected OM which  
had no implications for the operation of the OM.

Together with Line 0, the so-called MILOM (Fig. \ref{milom}), a 180 m instrumented line with one fully equipped storey, 
was deployed and connected at the same time. 
Its goal is to survey the environmental parameters and to test the calibration system for the future full detector with 
its laser beacons. 
A seismometer is linked by a 50 metre cable to the MILOM BSS and
on 30 August it has registered earthquake in Japan with an amplitude of 6.2 on the Richter scale. 
The BSS holds the SCM, a Laser Beacon and the acoustic positioning receiver-transmitter. 
The bottom storey (Bottom LCM) lies 100 m above and holds a LED Beacon, a Conductivity-Temperature probe (CT-Seabird), 
a light transmission meter (CSTAR) and a hydrophone acting as acoustic positioning receiver.
The storey in the middle (MLCM), located 15 m above the Bottom LCM and 50 m below the Top LCM holds 3 OMs and a 
sound velocimeter.
The Top LCM contains one OM, one LED beacon, the Acoustic Doppler Current Profiler (ADCP) and the so-called 
\textsl{Spy} hydrophone.

\section{Results from the MILOM}
The excellent angular resolution of ANTARES relies on the timing resolution and on an accurate knowledge 
of the OMs positions. 
With the MILOM, the Collaboration has been able to verify the required specification for the absolute clock timing ($\sim$1 ms), 
the relative timing of the OMs ($\sim$0.5 ns), and the required precision on the 3D position determination of the OMs ($\sim$10 cm).
The clock distribution system has been monitored by measuring the time delay in the 42 km long MEOC and 
between the Bottom and the Top LCMs. 
Results on the relative timing resolution of the OMs have been obtained by flashing the three OMs on the MLCM 
with high amplitude signals from the LED beacon in the Bottom LCM. 
The distributions of the time differences between the three OMs and the internal PMT of the 
LED beacon after subtraction of the light propagation offset have a $\sigma$ of the order of 0.5 ns 
(Fig. \ref{data} top).
The distribution of the time difference between two signals seen by 2 OMs in the MLCM is 0.75 ns as expected 
($\sqrt{2}\times0.5$ ns) (Fig. \ref{data} bottom).
Data from 1-dim acoustic positioning measurements have been analysed and the precision turned out to be 
less than 3 cm.  
The sound velocimeter records a value of about 1545 m/s, while the ADCP shows that there is a preferred 
current in the west-east direction and that the average current speed value is about 10 cm/s.
The water temperature is monitored in the JB and by the CT-Seabird. 
Both devices show that water temperature is between $13.2\;^\circ$C and $13.3\;^\circ$C. 
Light background due to $^{40}$K decays and bioluminescence activities  
is being continuosly monitored by the OMs on the MLCM. 
The levels and variations of single photon counting rates seen with the PSL \cite{Circella}
have been confirmed with the MILOM operation.
During June 2005, the background baseline rates were found to be between 90 and 140 kHz. 

\section{Conclusions} 
The operation of Line 0 and the MILOM has been a big step forward in the ANTARES 
project: the time resolution of about $0.5$ ns in the electronics has been demonstrated 
and the feasibility of the deployment and the ROV connection method 
has been proven. 
Given the experience acquired with the two test lines, Line 1 is now being assembled at 
CPPM in Marseille. It will be deployed in the winter '05-'06.
The full detector is expected to be operational by the end of 2007.

\section{References}


\medskip
\begin{thebibliography}{9}
\bibitem{Proposal} 
ANTARES Collaboration 1999 Proposal for a 0.1 km$^{2}$ detector, available at http://antares.in2p3.fr.
\bibitem{Transm} 
Aguilar J.A. et al. 2005 {\it Astropart. Phys.} {\bf 23} 131-155.
\bibitem{OM} 
Aguilar J.A. et al. 2002 {\it Nucl. Instr. and Methods} A {\bf 484}  369.
\bibitem{Sedim} 
Amram P. et al. 2003 {\it Astropart. Phys.} {\bf 19} 253-267.
\bibitem{MACRO} 
Ambrosio M. et al. 2001 {\it Astrophys. J.} {\bf 546} 1038-1054.
\bibitem{Circella} 
Circella M. et al. 2003 {\it proc. of the 28th ICRC, Tsukuba, Japan} 1529.
\end{thebibliography}
\end{document}